\documentclass[letterpaper]{article} 
\usepackage{aaai24}  
\usepackage{times}  
\usepackage{helvet}  
\usepackage{courier}  
\usepackage[hyphens]{url}  
\usepackage{graphicx} 
\usepackage{natbib}  
\usepackage{caption} 
\usepackage{algorithm}
\usepackage{algorithmic}
\usepackage{amsmath}
\usepackage{newfloat}
\usepackage{listings}
\DeclareCaptionStyle{ruled}{labelfont=normalfont,labelsep=colon,strut=off} 
\floatstyle{ruled}
\newfloat{listing}{tb}{lst}{}
\floatname{listing}{Listing}
\usepackage{multirow}
\usepackage{booktabs}

\title{Large Language Models are Not Stable Recommender Systems} 
\author {
    Tianhui Ma\textsuperscript{\rm 1},
    Yuan Cheng\textsuperscript{\rm 2},
    Hengshu Zhu\textsuperscript{\rm 2},
    Hui Xiong \textsuperscript{\rm 3}
}
\affiliations {
    \textsuperscript{\rm 1}University of Science and Technology of China\\
    \textsuperscript{\rm 2}Career Science Lab, BOSS Zhipin\\
    \textsuperscript{\rm 3}Hong Kong University of Science and Technology (Guangzhou)\\
    matianhui@mail.ustc.edu.cn,
    chengyuan02@kanzhun.com, zhuhengshu@gmail.com ,
 xionghui@ust.hk
}

\usepackage{bibentry}

\begin{document}

\maketitle

\begin{abstract}

With the significant successes of large language models (LLMs) in many natural language processing tasks, there is growing interest among researchers in exploring LLMs for novel recommender systems. However, we have observed that directly using LLMs as a recommender system is usually unstable due to its inherent position bias. To this end, we introduce exploratory research and find consistent patterns of positional bias in LLMs that influence the performance of recommendation across a range of scenarios. Then, we propose a Bayesian probabilistic framework, STELLA (\textbf{St}abl\textbf{e} \textbf{LL}M for Recommend\textbf{a}tion), which involves a two-stage pipeline. During the first probing stage,  we identify patterns in a transition matrix using a probing detection dataset. And in the second recommendation stage, a Bayesian strategy is employed to adjust the biased output of LLMs with an entropy indicator. Therefore, our framework can capitalize on existing pattern information to calibrate instability of LLMs, and enhance recommendation performance. Finally, extensive experiments clearly validate the effectiveness of our framework.

\end{abstract}

\section{Introduction}
Recommender systems have been considered as an important technology for facilitating a wide range of online services, such as news feed~\cite{wu2022feedrec}, video entertainment~\cite{papadamou2022just}, and display advertising \cite{display}. Traditional recommendation models primarily depend on historical user interaction behaviors (e.g., clicked item sequences)~\cite{DBLP:journals/corr/HidasiKBT15,DBLP:conf/icdm/KangM18}, and thus have limited capability to capture user preference in some complex contexts that require extensive knowledge \cite{DBLP:journals/tkde/GuoZQZXXH22}. With the significant successes of large language models (LLMs) in many Natural language processing (NLP) tasks, an increasing number of researchers have been showing interest in utilizing LLMs to develop generative recommender systems~\cite{gao2023chatrec,wang2023zeroshot,wang2023generative,dai2023uncovering,lin2023sparks,wu2023survey}.



\begin{figure}[t]
    \centering
    \includegraphics[width=0.9\columnwidth]{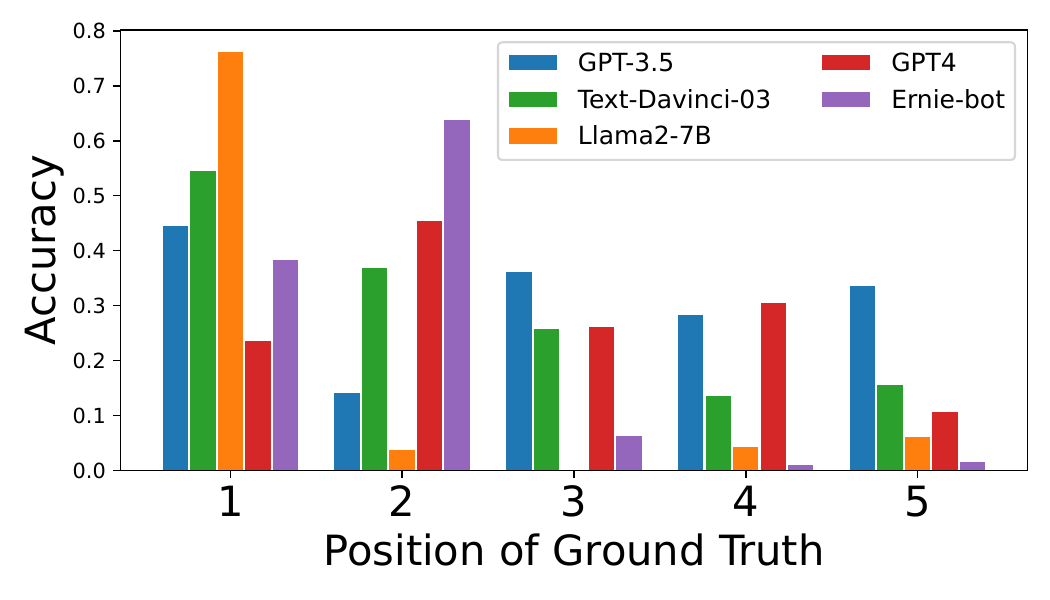}
    \caption{The recommendation performance of different LLMs on the Book dataset, when changing the position of ground truth items within the prompt. The results indicate that using LLMs as recommender system is unstable due to the inherent position bias.}
    \label{fig:one case}
\end{figure}


In recent studies, LLMs are exhibited to be highly sensitive to the prompt designs for many NLP tasks \cite{DBLP:conf/acl/LuBM0S22,DBLP:conf/icml/ZhaoWFK021, bowman2023eight, icl-survey, unfaithful-cot}. In particular, when directly using LLMs as a recommender system, the input items are usually transformed into the descriptions in prompt~\cite{hou2023large, liu2023chatgpt}, leading to unstable recommendation results that is sensitive to the order of the input candidate items. For example, as shown in Fig.\ref{fig:one case}, the recommendation performance will vary significantly when changing the position of ground truth items. Indeed, traditional recommender systems mainly focus on accuracy while ignoring the problem of variance~\cite{wu2023survey}, making them not sensitive to the order of input items. However, for using LLMs as recommender systems, it is important to consider both accuracy and variance (i.e., to reduce the variance while ensuring competitive recommendation accuracy), which is a challenging and under-explored task.

\begin{figure*}[t]
    \centering
    \includegraphics[width=0.90\textwidth]{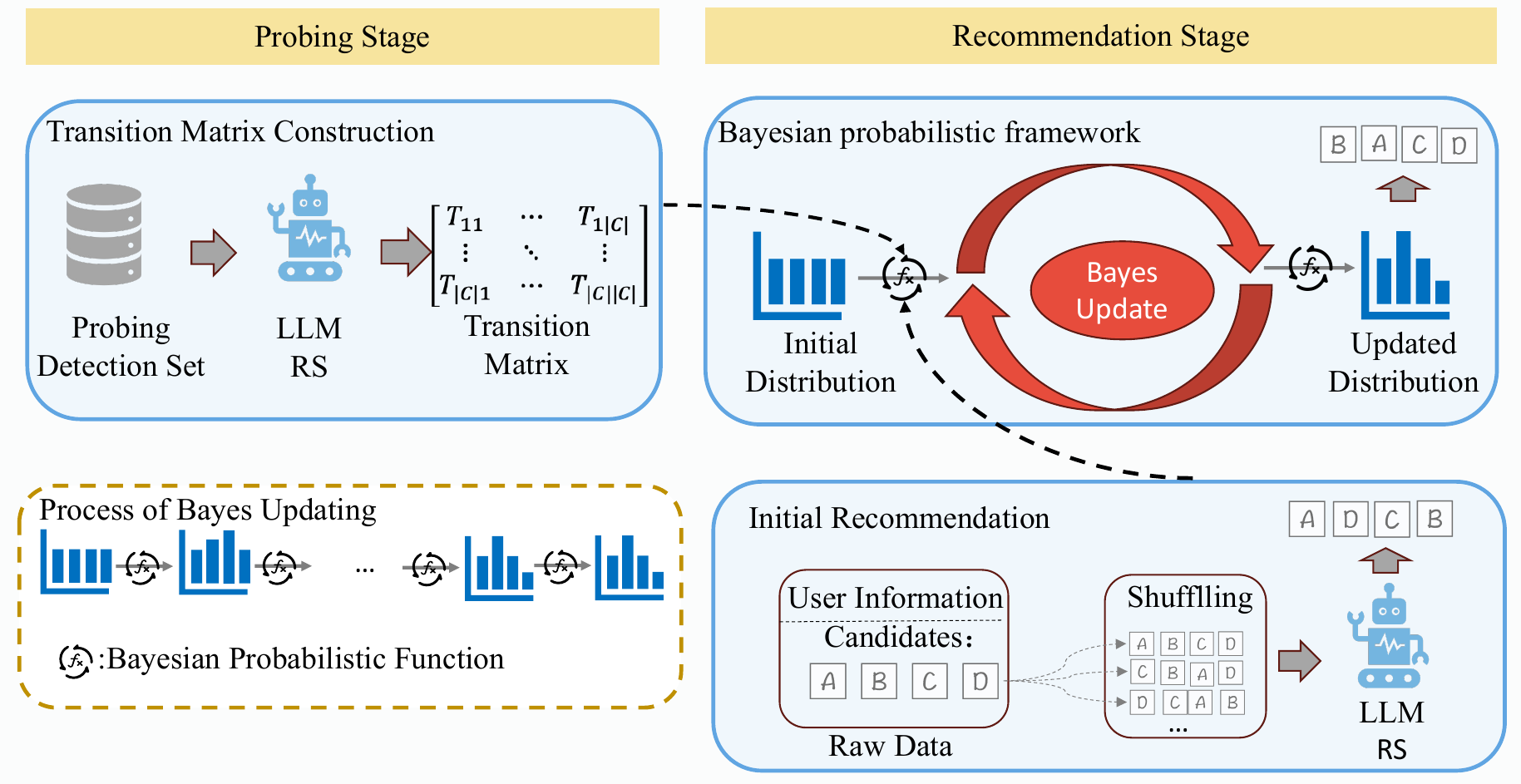}
    \caption{Overview of STELLA framework, which consists of a probing stage and a recommendation stage.}
    \label{fig:model}
\end{figure*}

To this end, in this paper, we introduce an exploratory
research on addressing the position bias of LLMs as recommender systems. Through an in-depth analysis, we first identify the consistent patterns of position bias that influence the recommendation's performance across a range of scenarios. Furthermore, we propose a Bayesian probabilistic framework, STELLA (\textbf{St}abl\textbf{e} \textbf{LL}M for Recommend\textbf{a}tion), which introduces a two-stage pipeline when using LLMs as recommender systems. As shown in Fig.\ref{fig:model}, STELLA consists of two interactive learning stages, namely the probing stage and the recommendation stage. The core idea is to introduce a probing dataset specifically designed for detecting position  bias patterns within LLMs. These patterns can then be employed to assess the confidence of recommendation outcomes, allowing for the calibration of biases in the recommender system. Specifically, during the probing stage, we identify patterns in a transition matrix using a probing detection dataset. In the recommendation stage, a Bayesian updating strategy is employed to adjust the biased output of LLMs, and a confidence indicator is introduced based on the output distribution entropy. Therefore, our framework can capitalize on existing pattern information to calibrate the instability of LLMs, and enhance recommendation performance through the confidence indicator.



Our contributions can be summarized as follows:
\begin{itemize}
    \item We identify consistent patterns related to position  bias specifically in the application of using LLMs as recommender systems, underscoring the necessity to address both variance and accuracy in this context.
    \item We first propose a novel Bayesian probabilistic framework, STELLA,  which introduces a two-stage pipeline when using LLMs as recommender system.
    \item We conduct extensive experiments that clearly validate the effectiveness of our framework, which can significantly reduce the variance and enhancing overall recommendation performance of LLMs.
\end{itemize}

\section{Related Work}
Generally, the related works of this paper can be grouped into two categories, namely LLM for recommendations and position bias in LLM.

\subsection{Large Language model for recommendations}
Large Language Models are emerging as powerful tools in the field of NLP ~\cite{DBLP:conf/nips/BrownMRSKDNSSAA20,DBLP:conf/nips/KojimaGRMI22,chowdhery2022palm} and have recently gained significant attention in the field of recommender systems. \citet{wu2023survey} provides a systematic review and analysis of existing LLMs for recommendation systems.

Recently, researchers have been exploring the potential of leveraging the powerful capabilities of LLM as a novel recommendation tool. \citet{liu2023chatgpt} and \citet{zhang2021language} systematically evaluate the performance of ChatGPT~\cite{chatgpt} on common recommendation tasks. Similarly, \citet{dai2023uncovering} conduct an empirical analysis of ChatGPT's recommendation abilities from three types of ranking: point-wise, pair-wise, and list-wise. \citet{hou2023large} introduce two prompting methods to improve the sequential recommendation ability of LLMs and have explored some basic problems in this new paradigm. 

However, it is necessary to note that the overall exploration in this field is still in the early stages. It cannot be negected that there is some inherent bias inside the LLMs as well as fairness issues related to sensitive attributes \cite{DBLP:journals/corr/abs-2304-03738}, which are influenced by the training data and the demographics of annotators. This leads to discussions and concerns regarding the challenges and risks of bias in LLMs for recommender systems. 


\subsection{Position Bias in Large Language Model}

Large Language Models (LLMs) are shown to exhibit position bias \cite{DBLP:conf/acl/LuBM0S22,DBLP:conf/icml/ZhaoWFK021, bowman2023eight, icl-survey, unfaithful-cot}, that is, an excessive dependency on the position of information position in prompts. 
Recent studies, such as \citet{wang2023large}, further explore the position bias of GPT-4 and ChatGPT when used as evaluators, showcasing a sensitivity to the sequence of candidate items. 
Moreover, \citet{qin2023large} find that simplifying listwise ranking tasks to pairwise comparisons can yield better performance. 
This problem also exists when applying LLMs for retrieval tasks. In the context of long-input retrieval, \citet{liu2023lost} find that language models oTable 3en struggle to use information in the middle of long input contexts, and that performance decreases as the input context becomes longer. 

However, it is important to note that the overall exploration of using LLMs as recommender systems is still in the early stages. 
The position bias problem in this new system needs to be explored systematically. 
\citet{hou2023large} investigate the capacity of LLMs to act as the ranking model for recommender systems, demonstrating that LLMs struggle to perceive the order of historical interactions and can be affected by position bias. 
Along this line of research, we systematically investigate this problem and the pattern underlying the phenomenon. We then propose a two-stage framework to address this problem.

\section{Instablity of LLMs as Recommender Systems}
In this section, we present the pipeline for using LLMs for sequential recommendation, along with a formal definition of related concepts. The design of the prompt is in Table \ref{table: prompt format}. Then, we show the patterns found within the phenomenon of position bias in LLM-based recommendation systems.
\begin{table}[t]
\caption{Prompt format}
\centering
\small
\label{table: prompt format}
\begin{tabular}{p{\columnwidth}} 
\toprule
Prompt \\
\midrule
You are a book recommendation system now. Please list the ranked recommendations. The output should be in the format of json, e.g.  \{``rank\_order":[``A", ``B", ``C", ``D", ``E"]\}.

Input: Here is the reading history of a user: Inferno, An Abundance of katherines, The Son, Joyland, The Guns at Last Light: The War in Western Europe, 1944-1945 (Liberation Trilogy).
The books on the candidate list are:

(A) No Easy Day: The Autobiography of a Navy Seal: The Firsthand Account of the Mission That Killed Osama Bin Laden,

(B) The Execution of Noa P. Singleton: A Novel,

(C) Allegiant,

(D) The Geography of Bliss: One Grump's Search for the Happiest Places in the World,

(E) Billy Lynn's Long HalTableime Walk: A Novel.

Output: \\
\bottomrule
\end{tabular}
\end{table}

\subsection{Notation and Problem Setting}
Our primary focus is on encoding traditional sequence recommendation user-item interactions into prompt formats, followed by subsequent decoding processes. The main component of the input is composed of three components:

\textbf{Task Description (I):} By explicitly providing instructions, we delineate the use of the model for recommendation systems.

\textbf{User Historical Behavior Description (H):} This represents the comprehensive historical interactions between the user and the item in a natural language format.
\begin{equation}
H=\text{Encoder}_{h}(h_1,h_2,\ldots,h_i),
\end{equation}
where \( h_i \) represents one piece of history record.

\textbf{Candidate Items (C):} This represents the whole set of candidate items to be ranked.
\begin{equation}
C=\text{Encoder}_{c}(c_1,c_2,\ldots,c_j),
\end{equation}
where \( c_j \) represents a candidate item in a natural language format, such as the title of the item.

Then we obtain the output:
\begin{equation}
\text{Output} = \text{LLM}(I,H,C).
\end{equation}
Upon analyzing the results, we discern the recommendation ranking results of the large model:
\begin{equation}
Y=y_1,y_2,\ldots,y_j = \text{Decoder}(\text{Output}).
\end{equation}
The output items then undergo a basic legality check, and invalid answers are excluded.

Specifically, the conversion of the original candidate set into natural language introduces sequential information, affecting the recommendation results of the system. The sequence of items to be ranked introduces diversity into \( C \). Considering all the permutations, there are \( j! \) ways to arrange the sequence \( \{c_1,c_2,\ldots,c_j\} \). Formally, it can be represented as follows:
\begin{equation}
C \in [C_1,\ldots,C_k],
\end{equation}
where \( k=j! \). Correspondingly,
\begin{equation}
Y \in [Y_1,\ldots,Y_k],
\end{equation}
where \( k=j! \). Given the evaluation metrics, we obtain a performance score \( S_i \) for each ranked output \( Y_i \). We calculate the mean value and the variance of \( S_i \), which represent the capability and stability of the system, respectively.

\subsection{Position Bias}
In this section, we explore the characteristics of ChatGPT (GPT-3.5-turbo) when using it as a recommender system. The investigation primarily focuses on the sensitivity and patterns of the model in four aspects: the prompt template, candidate size, attention to context, and permutation.

\subsubsection{Prompt Template.}
\begin{figure}[t]
    \centering
    \includegraphics[width=0.9\columnwidth]{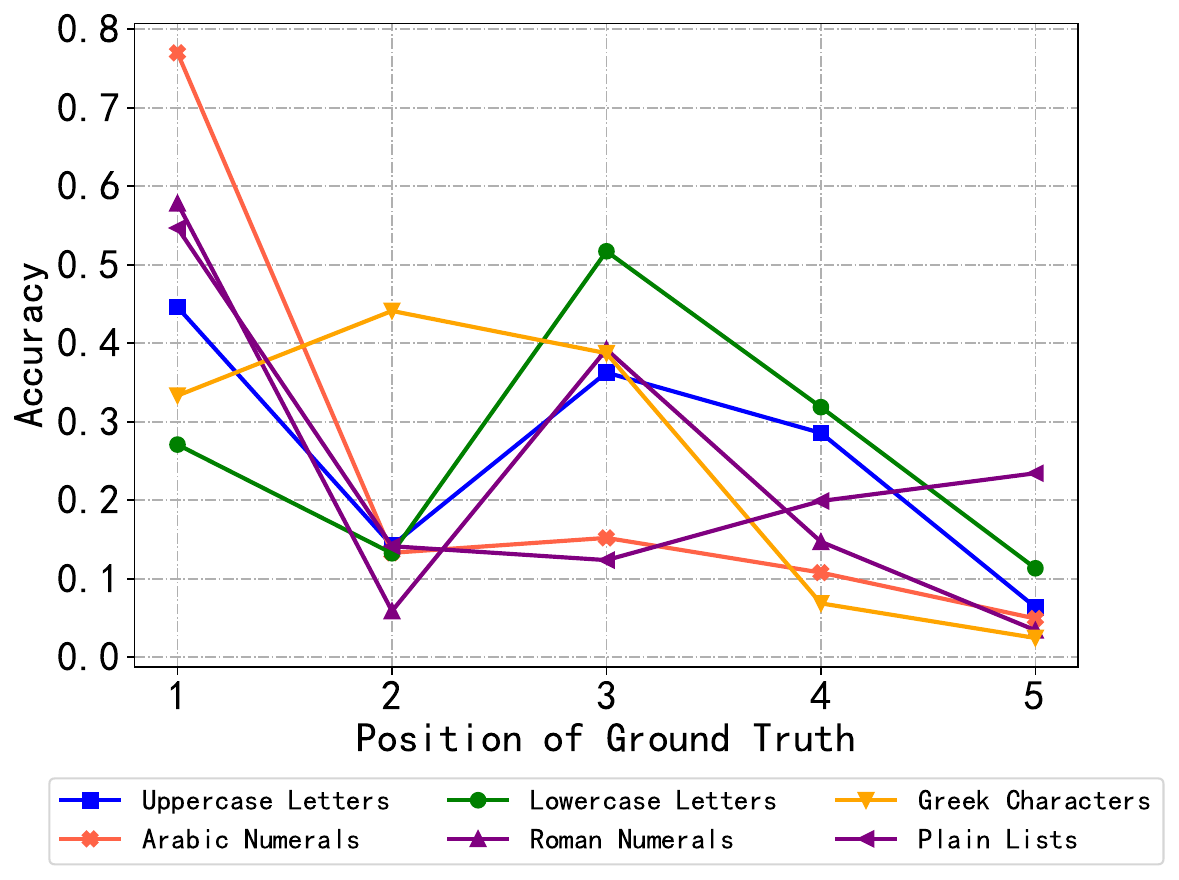}
    \caption{Model performance with respect to different prompt templates. }
    \label{fig:templates_acc}
\end{figure}

For the prompt template, we test various variations, such as removing candidate labels \ (e.g. A/B/C) or replacing candidate labels in Table \ref{table: prompt format}  with Arabic numerals, lowercase letters, Greek characters, Roman numerals, or plain lists (e.g. Candidate 1/Candidate 2). As illustrated in Fig.\ref{fig:templates_acc}, different prompt templates lead to distinct trends in the accuracy of the model's output. However, all of them reveal the issue of position bias, where the model's recommendation ability is influenced by the different input sequences for the candidate items.
\subsubsection{Permutation.} 
\begin{figure}[t]
    \centering
    \includegraphics[width=0.75\columnwidth]{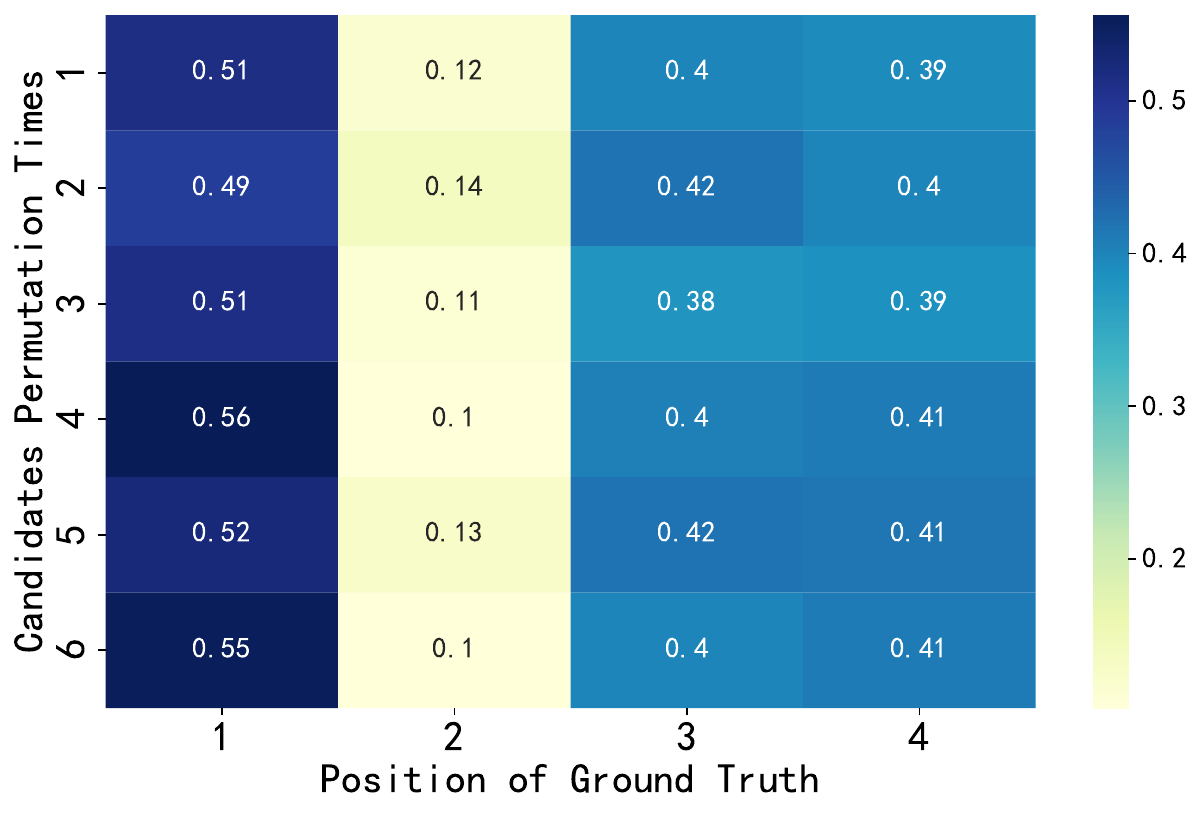}
    \caption{For each position of ground truth, we permutate the negative samples.}
    \label{fig:heat_map}
\end{figure}

Figure~\ref{fig:heat_map} shows the situation with four candidate items. The x-axis represents the position of the ground truth in the candidate items, and the y-axis represents the permutation situation of the negative samples. There are three negative samples, corresponding to six permutations. It shows the changes in accuracy under these combination scenarios. For the same x-value column, the accuracy variation is small, while there is significant variation for the same y-value row. This indicates that the model is less sensitive to the arrangement order of the negative samples compared with the position of the ground truth.

\subsubsection{Position Sensitivity to Context.}

\begin{figure*}[t]
    \centering
    \includegraphics[width=0.31\textwidth]{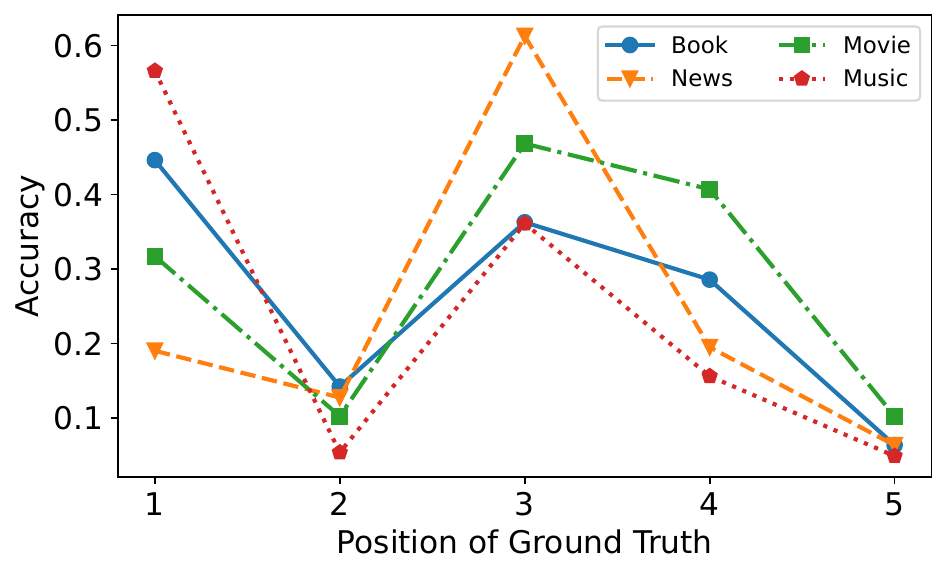}
    \includegraphics[width=0.31\textwidth]{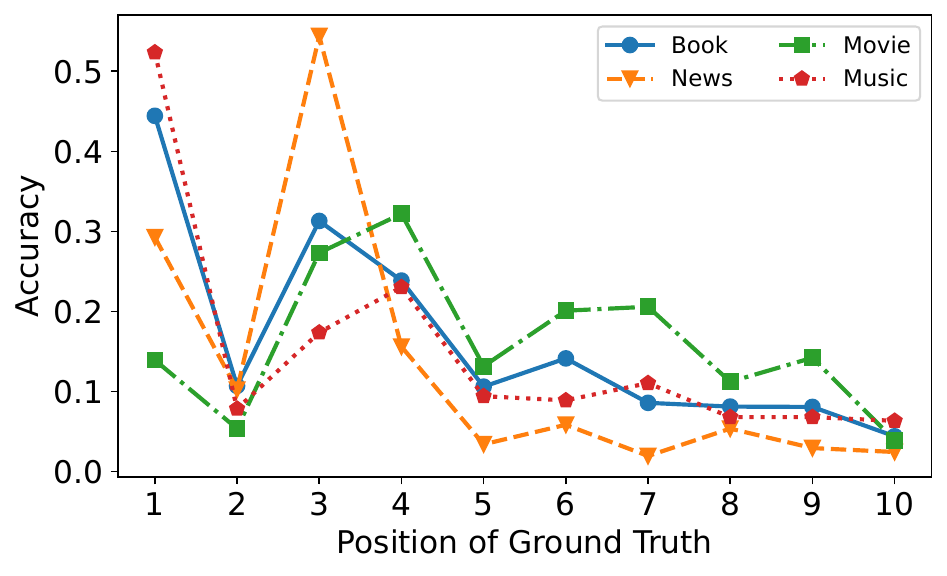}
    \includegraphics[width=0.31\textwidth]{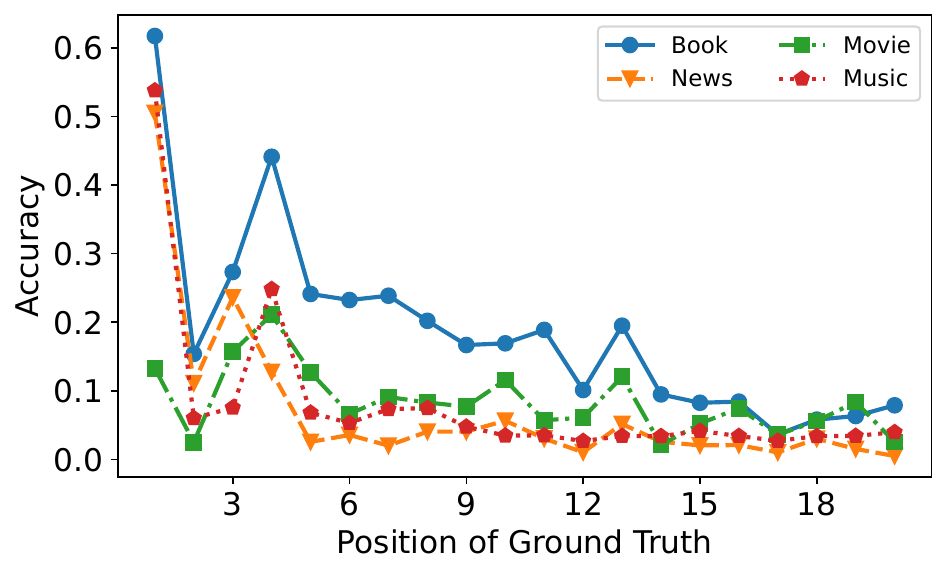}
    
    \caption{Accuracy across different candidate set sizes on four datasets. The results shows position sensitivity to the context.}
    \label{fig:all_length_acc}
\end{figure*}

From Fig.\ref{fig:all_length_acc}, we observe significant variations in the  recommendation performance of the model that are related to the context of candidate items. This position sensitivity to context manifests in two significant patterns, as elaborated below.

(1)Limited Sensitivity of the model to the Context of Candidate Items:
As shown in Fig.\ref{fig:all_length_acc},when the length of the candidate set is 20, the corresponding average recommendation accuracy is as low as 0.05. Specifically, in the Book dataset, the output resembles random results aTableer the 15th item, while in the other three datasets, the output approximates random results after just five items.
This pattern suggests that the model can effectively focus on a limited number of candidate items when making recommendations. When this number is small, the recommendations are more accurate. However, as the number of candidate items increases,  the recommendations become increasingly inaccurate, resembling random selections. This underlines the need for carefully evaluating and selecting an appropriate candidate set length in practice to avoid significant degradation in recommendation quality.

(2) Model Struggles to Accurately Identify the Last Position:
As is evident from Fig.\ref{fig:all_length_acc}, across different candidate set sizes, when the true value is placed at the end of the candidate items, the model consistently exhibits almost the worst results.

\subsubsection{Candidate Size.}
\begin{figure}[t]
    \centering
    \includegraphics[width=0.95\columnwidth]{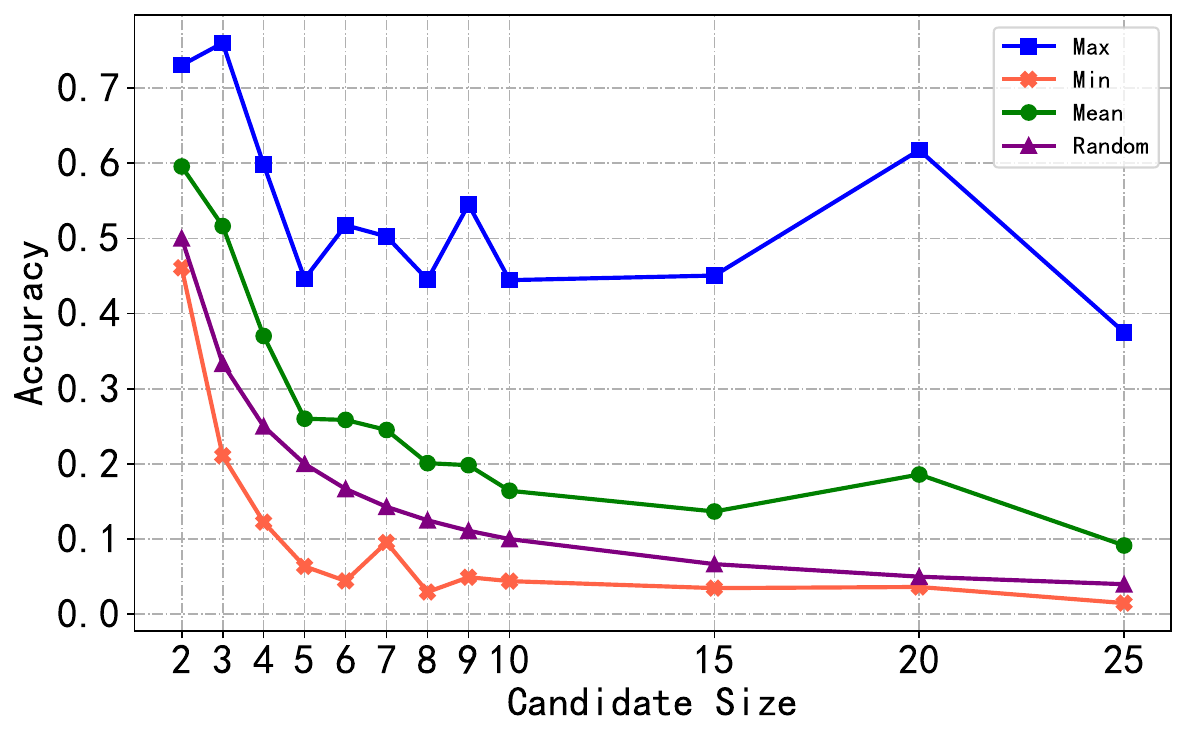}
    \caption{Model performance with different candidate sizes.}
    \label{fig:candsvsacc}
\end{figure}

Figure~\ref{fig:candsvsacc} shows the performance of the model substantially decreases as the size of the candidate set increases. To investigate this, we use a fixed prompt template and choose different sizes of candidate items. For each size, we place the ground truth at all corresponding positions to obtain all possible accuracies. We display the maximum, minimum, and average values. Additionally, we present the results of random recommendations as a comparison baseline.
Figure \ref{fig:candsvsacc} shows that the model's accuracy changes a lot when the size of the candidate set increases from 2 to 25. For example, when the candidate set size is 2, the difference between the highest and lowest accuracy values is 0.2. This difference increases to 0.4 or more when the candidate set size is larger than 3. The model's recommendations are consistently better than random recommendations. In Figure \ref{fig:candsvsacc}, the average accuracy of the model's recommendations is always higher than the accuracy of random recommendations.

However, this better performance decreases as the candidate set size gets larger. For instance, the average accuracy is 0.6 for a candidate set size of 2, but it drops to 0.1 for a size of 25. The improvement over random recommendations also changes, being 0.1 for a size of 2 and 0.15 for a size of 3, and it varies at other sizes.

These findings highlight the importance of choosing the right candidate set size when using large models for recommendation systems, as this choice has a big effect on how well the recommendations work.

\section{Calibrating the Position Bias}
Given the sequential histories with candidates, our task is to rank these candidates using Large Language Models (LLMs) as a recommender system. In this section, we introduce the technical details of our Bayesian probabilistic framework, STELLA (\textbf{St}abl\textbf{e} \textbf{LL}M for Recommend\textbf{a}tion), which employs a two-stage pipeline when using LLMs as a recommender system. During the probing stage, we identify patterns in a transition matrix using a probing detection dataset. In the recommendation stage, a Bayesian strategy is employed to adjust the biased output of LLMs with an entropy indicator. The overview of this proposed framework is illustrated in Figure \ref{fig:model}.

\subsection{Probing Stage}
Given the user histories, a probing set is created and integrated into the model. By analyzing the output, we obtain a transition matrix that reflects the position bias of the model.

First, the probing set is created in the same manner as the true test set, using only histories, and leaving one as ground truth while extracting several previous histories. The candidate set consists of the ground truth and several negative samples. Then, for each piece of probing data, we permute the position of the ground truth, creating ensemble data that makes up the final probing dataset. Finally, by analyzing the output of the model while inferring from the probing set, we derive the transition matrix $T$, which reflects the transition from the original ground truth position to the predicted ones. Formally, \( T \) is defined as follows
\begin{equation}
    T = \begin{bmatrix} T_{11} & T_{12} & \ldots & T_{1|C|} \\ T_{21} & T_{22} & \ldots & T_{2|C|} \\ \vdots & \vdots & \ddots & \vdots \\ T_{|C|1} & T_{|C|2} & \ldots & T_{|C||C|} \end{bmatrix},
\end{equation}
where
\( T_{ij} \) is the conditional probability that the model ranks an item at position \( j \) first while ground truth locating at position \( i \) and $|C|$ is the number of candidate positions.

As discussed in the pattern-revealing section, there are position biases in the system. The probing dataset is capable of capturing some user preferences from the histories. Through the ensemble construction and analysis of the model's output pattern, consistent position bias information can be transferred from this training stage to the inference stage. The format of the transition information in relation to the transition matrix will be further discussed in the section of experiments. 

\subsection{Recommendation Stage}
Based on the prediction of the model during the inference period, we post-process it by utilizing the transition matrix from the probing stage to calibrate the position bias inside the direct output of the model.

We apply Bayesian updating to correct the  biased output of the model for ranking candidates, using entropy as an indicator to infer the final result. The procedure is as follows:

\begin{itemize}
    \item \textbf{Initialization}: In the recommendation stage, we begin by initializing the prior probability vector \( \mathbf{p} \) of the ground truth's position to a uniform distribution.
    
    \item \textbf{Bayesian Updating}: Next, we utilize the  prediction result \( \mathbf{y} \) of the model and the transition matrix \( \mathbf{T} \) obtained from the probing phase to compute the posterior probability distribution \( \mathbf{p}_{\text{new}} \) via Bayesian updating:
    \begin{equation}
    \label{equ:bayes}
            \mathbf{p}_{\text{new}} = \frac{\mathbf{p} \odot \mathbf{T}_{y}}{\sum_{i=1}^n (\mathbf{p} \odot \mathbf{T}_{y})_i}.
    \end{equation}
    where \( \odot \) represents element-wise multiplication, and \( \mathbf{T}_{y} \) is the \( y \)-th column of \( \mathbf{T} \), representing the conditional probability distribution when the model's predicted first item is at position \( y \).

    \item \textbf{Entropy Calculation}: Subsequently, we determine the entropy \( H(\mathbf{p}_{\text{new}}) \) of the posterior probability distribution \( \mathbf{p}_{\text{new}} \):
    \begin{equation}
    \label{equ:entropy}
            H(\mathbf{p}_{\text{new}}) = -\sum_{i=1}^n p_{\text{new},i} \log p_{\text{new},i}.
    \end{equation}

    \item \textbf{Convergence and Result}: When the entropy changes slightly for several consecutive iterations, or the maximum number of iterations is reached, the algorithm halts. The final recommendation result is the ranking of the candidate projects corresponding to the posterior probability distribution \( \mathbf{p}_{\text{new}} \) with the smallest entropy.
\end{itemize}

By implementing this process, Bayesian inference seamlessly combines observed data and prior information, and it offers a more precise and confidence-informed prediction of the ground truth position. It thus aids in calibrating the position bias present in the direct output of the model, making it a practical solution for applications using large language models in recommendation systems.

As discussed in the previous section, using large language model as recommender systems exhibits consistent position bias, which prompts us to explore the distribution of the position of the ground truth. Solving this problem fundamentally is quite challenging due to the high cost of pretraining large language models. We argue that the post-process technique is necessary when applying large language models in our case of recommendation. Our framework is low-resource and intuitive. We aim to calibrate the position bias in this system based on the detected pattern and the direct output of the model. This aligns with Bayesian update theory, which seeks to calibrate biased priors based on old likelihoods and new evidence. The latter experiments section demonstrates the effectiveness of our framework.
\begin{algorithm}
\caption{Probability Based Post-processing Algorithm}
\label{alg:algorithm}
\textbf{Input:} Prior probability distribution \( \boldsymbol{p} \), Model's prediction result \( \boldsymbol{y} \), Transition matrix \( \boldsymbol{T} \) \\
\textbf{Parameter:} Maximum iteration number \(N\) \\
\textbf{Output:} Final Ranked result \(\hat{y}\), Entropy records \(E\) \\
\begin{algorithmic}[1] 
\STATE Initialize prior probability distribution \( \boldsymbol{p} \) as a uniform distribution
\FOR{iteration \(i = 1\) to \(N\)}
    \STATE Update posterior distribution using Equation \ref{equ:bayes}
    \STATE Calculate entropy \( e \) using Equation \ref{equ:entropy}, and save the record in \(E\)
    \IF {Entropy \(E\) converges}
        \STATE \textbf{return} \(\hat{y}\) corresponding to minimum \(e\)
    \ENDIF
\ENDFOR
\IF {Maximum iterations \(N\) reached and \(E\) has not converged}
    \STATE \textbf{return} \(\hat{y}\) corresponding to minimum \(e\) in \(E\)
\ENDIF
\end{algorithmic}
\end{algorithm}


\section{Experiments}
In this section, we use ChatGPT (GPT-3.5-turbo) as the LLM,  which has been widely used in recommendation system research, to perform the experiments for validating the effectiveness of STELLA.

\subsection{Datasets}
To more profoundly delve into the diverse capabilities of LLM in personalized recommendations, we conduct evaluations on four datasets encompassing a variety of domains.

\begin{itemize}
    \item \textbf{Movies:} We utilized the widely adopted MovieLens-1M dataset \cite{DBLP:journals/tiis/HarperK16}, encompassing one million user ratings for films.
    \item \textbf{Books:} We drew from the ``Books'' subset of the Amazon dataset \cite{DBLP:conf/www/HeM16}, containing user ratings for an array of literary works.
    \item \textbf{Music:} For the music domain, we utilized the ``CDs \& Vinyl'' from Amazon dataset \cite{DBLP:conf/www/HeM16}.
    \item \textbf{News:} The MIND-small dataset served as the benchmark for our exploration in the news domain \cite{DBLP:conf/acl/WuQCWQLLXGWZ20}.
\end{itemize}

Aligning with conventional practices \cite{DBLP:conf/www/HeLZNHC17,DBLP:conf/cikm/MaoZWDDXH21,DBLP:conf/cikm/Xu0CDW22}, for the movie, book, and music datasets, the feedback data with ratings higher than 3 are treated as positive samples, while all the other ratings as negative \cite{dai2023uncovering}. For the news dataset, we utilize the original sample labels. In our experiments, the title of the item is used as the description within the historical behaviors as well as the candidate items. Due to substantial cost constraints, we randomly selected 200 users from each dataset for experiments.
\subsection{Evaluation}
Following existing works \cite{DBLP:conf/icdm/KangM18,DBLP:conf/kdd/HouMZLDW22}, we apply the leave-one-out strategy for evaluation. For each historical interaction sequence, the last item is treated as the ground truth for evaluation.  For all experiments, the positive item are paired with randomly sampled negative items to consist a candidate item list. We use accuracy as ranking metrics for performance evaluation. 

\subsection{Implementation Details}
We can simply obtain probing set from training dataset. In this work we use the last $m$ historical behaviors of each user before the evaluation time step to form a probing set. In this work, $m$ is referred to as length of ensemble steps and $m$ is set to 5 in our experiments. For each scenario, after permuting the position of the ground truth within the candidates of the probing set, we can calculate the distribution of predicted true values and construct the corresponding transition matrix. Next, we initiate the Bayesian updating process. During each update, we shuffle the order of the input data and set the maximum sampling step as 10. We stop the updates either convergence or reaching the maximum sampling step. After stopping the updates, we use the results corresponding to the minimum entropy value among the updated results as the final output.

Our framework, STELLA, performs post-processing on the model's recommended results. Due to the instablity of LLMs, the raw recommendation output varies significantly. The main baseline in this experiment is the Bootstrapping strategy proposed by \cite{hou2023large} . The strategy includes a statistical inference process through permutation tests and an aggregation process for multiple votes, referred to as the Borda count method \cite{DBLP:journals/scw/Emerson13}. Specifically, scores are assigned to each item based on their ranking position, with items ranked higher receiving higher scores. For example, for candidate items A, B, and C, if the ranked result is B, C, and A, the first position receives 3 scores (in this case, B), while items C and A receive 2 scores and 1 score, respectively. By aggregating multiple ranked results, we rank the items based on their final total scores. 

We also use the Borda count method in STELLA. In our experiments, the aggregation parameter is set to 3, and for the baseline Bootstrapping, the results are obtained by aggregating 3 outputs. For STELLA, the lowest 3 ranked outputs corresponding to the entropy indicator are aggregated to produce the final result.


\subsection{Main Results}

We evaluate the effectivness of our framework. Table \ref{exp_main} shows the results of the all of the datasets, corresponding to raw output, Bootstrapping and STELLA, respectively.

\subsubsection{Stablity} The raw outputs from the LLM varies significantly, with the recommendation accuracy in a large range. In Table 2, we calculate the mean value and standard deviation of the raw output. From the standard deviation, we can observe the instability of the raw output. For instance, the accuracy on the Book dataset bottoms out at 0.2118, approximating a random outcome, yet it can peak at 0.3713. In contrast, STELLA remains consistent and stable due to its Bayesian updating framework.
    
\subsubsection{Accuracy} We employ accuracy as the metric to assess the effectiveness of recommendations. While Bootstrapping and our STELLA exhibit greater stability compared to the raw outputs, it is noted that the Bootstrapping often fails to surpass the average value of the raw output. However, our proposed STELLA approach consistently outperforms both Bootstrapping and average of raw output by a significant improvement of over 15\% across all four datasets.

\begin{table}[t]
\centering
\caption{Results of the experiments on four datasets.}
\label{exp_main}
\begin{tabular}{l*{3}{c}}
\toprule
& Raw Output & Bootstrapping & STELLA \\
\midrule
Book & \(0.2915_{\pm 0.0798}\) & 0.2647 & \textbf{0.3235}  \\
Movie & \(0.2740_{\pm 0.0593}\) & 0.2537 & \textbf{0.2976}  \\
Music & \(0.2500_{\pm 0.0300}\) & 0.2650 & \textbf{0.3000}  \\
News & \(0.2610_{\pm 0.0219}\) & 0.2341 & \textbf{0.2732} \\
\bottomrule
\end{tabular}
\end{table}


\subsection*{Ablation Study}
In this work, calculating the transition matrix based on the probing detection set is the key technique. We examine the length of ensemble steps to investigate the influence of the transition matrix on the effectiveness of recommendations.

\subsubsection{Transition Matrix.}

Within our STELLA, the transition matrix is utilized during the Bayes update process. To test the performance in absence of the transition matrix, we substitute it with a random uniform matrix, which is equivalent to the assumption without position bias.  As indicated in the Table \ref{exp_aba}, we can find that the accuracy on all the datasets is significantly descreased. This again underscores the necessity of taking position bias into consideration.

\begin{table}[t]
\centering
    \caption{ Ablation study about Transition Matrix.}
    \label{exp_aba}
    \begin{tabular}{l*{4}{c}}
    \toprule
    Method & Book & Movie & Music & News \\
    \midrule
    STELLA& \textbf{0.3235} & \textbf{0.2976} & \textbf{0.3000} & \textbf{0.2732} \\
    $W/O$ TM & 0.2696 & 0.2439 & 0.2450 & 0.2390 \\
    \bottomrule
    \end{tabular}
\end{table}

\subsubsection{Size of Probing Detection Set.}
\begin{figure}[t]
    \centering
    \includegraphics[width=0.95\columnwidth]{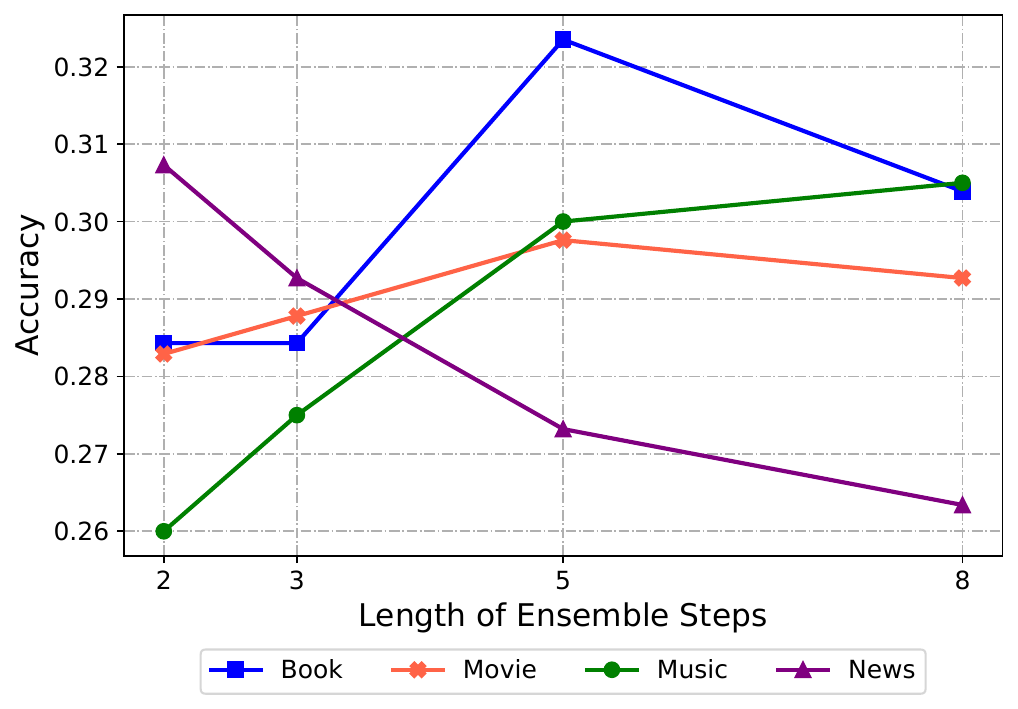}
    \caption{Ablation Study about Length of Ensemble Steps for Probing Detection Set.}
    \label{fig:abalation}
\end{figure}

We test different lengths while ensembling the probing detection set. As shown in Fig.\ref{fig:abalation}, among the Book, Movie, and Music datasets, the accuracy of our framework, STELLA, improves as the length of ensemble steps increases. However, the opposite is observed in the News dataset. This discrepancy may be attributed to the inherent time sensitivity of news recommendations, which differs from the other three scenarios. As the number of training time steps increases, the transition matrix obtained from the probing detection set does not more accurately reflect user preferences. 
However, better results are achieved when the length of ensemble steps is set to 5. 



\subsection{Discussion}
In this work, we conduct experiments using ChatGPT as a Large Language Model. When using relatively smaller-scale Large Language Models, we observe that their outputs tend to either mirror the inputs or produce invalid outputs. This behavior may be attributed to the demanding capability requirements of the recommendation task. Our proposed framework, STELLA, focuses solely on exploring behavioral patterns in the Large Language Model's inference for the recommendation scenario, rather than on training or retraining the model. As a result, the associated costs remain relatively low.

\section{Concluding Remarks}
In this paper, we study the instability problem that arises when using Large Language Models as recommender systems. Through detailed analysis, we identify consistent patterns of bias that influence performance across a range of scenarios. Based on these characteristics, we develop the intuition to extract consistent patterns of instability and use this information to calibrate the biased output each time the system makes an inference. To address this issue, we propose a Bayesian probabilistic framework, named STELLA, which introduces a two-stage pipeline for using Large Language Models as recommender systems. During the probing stage, we identify patterns in a transition matrix using a probing detection dataset. In the recommendation stage, we employ a Bayesian updating strategy to adjust the biased output of Large Language Models, and we introduce a confidence indicator based on the entropy of the output distribution. As a result, our framework is able to capitalize on existing pattern information to calibrate the instability of Large Language Models and enhance recommendation performance through the confidence indicator. Finally, extensive experiments clearly validate the effectiveness of our framework.

\bibliography{aaai24}

\begin{thebibliography}{36}
\providecommand{\natexlab}[1]{#1}

\bibitem[{Bowman(2023)}]{bowman2023eight}
Bowman, S.~R. 2023.
\newblock Eight things to know about large language models.
\newblock \emph{arXiv preprint arXiv:2304.00612}.

\bibitem[{Brown et~al.(2020)Brown, Mann, Ryder, Subbiah, Kaplan, Dhariwal,
  Neelakantan, Shyam, Sastry, Askell, Agarwal, Herbert{-}Voss, Krueger,
  Henighan, Child, Ramesh, Ziegler, Wu, Winter, Hesse, Chen, Sigler, Litwin,
  Gray, Chess, Clark, Berner, McCandlish, Radford, Sutskever, and
  Amodei}]{DBLP:conf/nips/BrownMRSKDNSSAA20}
Brown, T.~B.; Mann, B.; Ryder, N.; Subbiah, M.; Kaplan, J.; Dhariwal, P.;
  Neelakantan, A.; Shyam, P.; Sastry, G.; Askell, A.; Agarwal, S.;
  Herbert{-}Voss, A.; Krueger, G.; Henighan, T.; Child, R.; Ramesh, A.;
  Ziegler, D.~M.; Wu, J.; Winter, C.; Hesse, C.; Chen, M.; Sigler, E.; Litwin,
  M.; Gray, S.; Chess, B.; Clark, J.; Berner, C.; McCandlish, S.; Radford, A.;
  Sutskever, I.; and Amodei, D. 2020.
\newblock Language Models are Few-Shot Learners.
\newblock In \emph{NeurIPS}.

\bibitem[{Chen et~al.(2016)Chen, Sun, Li, Lu, and Hua}]{display}
Chen, J.; Sun, B.; Li, H.; Lu, H.; and Hua, X.-S. 2016.
\newblock Deep CTR Prediction in Display Advertising.
\newblock In \emph{Proceedings of the 24th ACM International Conference on
  Multimedia}, MM '16, 811–820. New York, NY, USA: Association for Computing
  Machinery.
\newblock ISBN 9781450336031.

\bibitem[{Chowdhery et~al.(2022)Chowdhery, Narang, Devlin, Bosma, Mishra,
  Roberts, Barham, Chung, Sutton, Gehrmann et~al.}]{chowdhery2022palm}
Chowdhery, A.; Narang, S.; Devlin, J.; Bosma, M.; Mishra, G.; Roberts, A.;
  Barham, P.; Chung, H.~W.; Sutton, C.; Gehrmann, S.; et~al. 2022.
\newblock Palm: Scaling language modeling with pathways.
\newblock \emph{arXiv preprint arXiv:2204.02311}.

\bibitem[{Dai et~al.(2023)Dai, Shao, Zhao, Yu, Si, Xu, Sun, Zhang, and
  Xu}]{dai2023uncovering}
Dai, S.; Shao, N.; Zhao, H.; Yu, W.; Si, Z.; Xu, C.; Sun, Z.; Zhang, X.; and
  Xu, J. 2023.
\newblock Uncovering ChatGPT's Capabilities in Recommender Systems.
\newblock arXiv:2305.02182.

\bibitem[{Dong et~al.(2022)Dong, Li, Dai, Zheng, Wu, Chang, Sun, Xu, and
  Sui}]{icl-survey}
Dong, Q.; Li, L.; Dai, D.; Zheng, C.; Wu, Z.; Chang, B.; Sun, X.; Xu, J.; and
  Sui, Z. 2022.
\newblock A Survey for In-context Learning.
\newblock \emph{arXiv preprint arXiv:2301.00234}.

\bibitem[{Emerson(2013)}]{DBLP:journals/scw/Emerson13}
Emerson, P. 2013.
\newblock The original Borda count and partial voting.
\newblock \emph{Soc. Choice Welf.}, 40(2): 353--358.

\bibitem[{Ferrara(2023)}]{DBLP:journals/corr/abs-2304-03738}
Ferrara, E. 2023.
\newblock Should ChatGPT be Biased? Challenges and Risks of Bias in Large
  Language Models.
\newblock \emph{CoRR}, abs/2304.03738.

\bibitem[{Gao et~al.(2023)Gao, Sheng, Xiang, Xiong, Wang, and
  Zhang}]{gao2023chatrec}
Gao, Y.; Sheng, T.; Xiang, Y.; Xiong, Y.; Wang, H.; and Zhang, J. 2023.
\newblock Chat-REC: Towards Interactive and Explainable LLMs-Augmented
  Recommender System.
\newblock arXiv:2303.14524.

\bibitem[{Guo et~al.(2022)Guo, Zhuang, Qin, Zhu, Xie, Xiong, and
  He}]{DBLP:journals/tkde/GuoZQZXXH22}
Guo, Q.; Zhuang, F.; Qin, C.; Zhu, H.; Xie, X.; Xiong, H.; and He, Q. 2022.
\newblock A Survey on Knowledge Graph-Based Recommender Systems.
\newblock \emph{{IEEE} Trans. Knowl. Data Eng.}, 34(8): 3549--3568.

\bibitem[{Harper and Konstan(2016)}]{DBLP:journals/tiis/HarperK16}
Harper, F.~M.; and Konstan, J.~A. 2016.
\newblock The MovieLens Datasets: History and Context.
\newblock \emph{{ACM} Trans. Interact. Intell. Syst.}, 5(4): 19:1--19:19.

\bibitem[{He and McAuley(2016)}]{DBLP:conf/www/HeM16}
He, R.; and McAuley, J.~J. 2016.
\newblock Ups and Downs: Modeling the Visual Evolution of Fashion Trends with
  One-Class Collaborative Filtering.
\newblock In \emph{{WWW}}, 507--517. {ACM}.

\bibitem[{He et~al.(2017)He, Liao, Zhang, Nie, Hu, and
  Chua}]{DBLP:conf/www/HeLZNHC17}
He, X.; Liao, L.; Zhang, H.; Nie, L.; Hu, X.; and Chua, T. 2017.
\newblock Neural Collaborative Filtering.
\newblock In \emph{{WWW}}, 173--182. {ACM}.

\bibitem[{Hidasi et~al.(2016)Hidasi, Karatzoglou, Baltrunas, and
  Tikk}]{DBLP:journals/corr/HidasiKBT15}
Hidasi, B.; Karatzoglou, A.; Baltrunas, L.; and Tikk, D. 2016.
\newblock Session-based Recommendations with Recurrent Neural Networks.
\newblock In \emph{{ICLR} (Poster)}.

\bibitem[{Hou et~al.(2022)Hou, Mu, Zhao, Li, Ding, and
  Wen}]{DBLP:conf/kdd/HouMZLDW22}
Hou, Y.; Mu, S.; Zhao, W.~X.; Li, Y.; Ding, B.; and Wen, J. 2022.
\newblock Towards Universal Sequence Representation Learning for Recommender
  Systems.
\newblock In \emph{{KDD}}, 585--593. {ACM}.

\bibitem[{Hou et~al.(2023)Hou, Zhang, Lin, Lu, Xie, McAuley, and
  Zhao}]{hou2023large}
Hou, Y.; Zhang, J.; Lin, Z.; Lu, H.; Xie, R.; McAuley, J.; and Zhao, W.~X.
  2023.
\newblock Large Language Models are Zero-Shot Rankers for Recommender Systems.
\newblock arXiv:2305.08845.

\bibitem[{Kang and McAuley(2018)}]{DBLP:conf/icdm/KangM18}
Kang, W.; and McAuley, J.~J. 2018.
\newblock Self-Attentive Sequential Recommendation.
\newblock In \emph{{ICDM}}, 197--206. {IEEE} Computer Society.

\bibitem[{Kojima et~al.(2022)Kojima, Gu, Reid, Matsuo, and
  Iwasawa}]{DBLP:conf/nips/KojimaGRMI22}
Kojima, T.; Gu, S.~S.; Reid, M.; Matsuo, Y.; and Iwasawa, Y. 2022.
\newblock Large Language Models are Zero-Shot Reasoners.
\newblock In \emph{NeurIPS}.

\bibitem[{Lin and Zhang(2023)}]{lin2023sparks}
Lin, G.; and Zhang, Y. 2023.
\newblock Sparks of Artificial General Recommender (AGR): Early Experiments
  with ChatGPT.
\newblock arXiv:2305.04518.

\bibitem[{Liu et~al.(2023{\natexlab{a}})Liu, Liu, Zhou, Lv, Zhou, and
  Zhang}]{liu2023chatgpt}
Liu, J.; Liu, C.; Zhou, P.; Lv, R.; Zhou, K.; and Zhang, Y. 2023{\natexlab{a}}.
\newblock Is ChatGPT a Good Recommender? A Preliminary Study.
\newblock arXiv:2304.10149.

\bibitem[{Liu et~al.(2023{\natexlab{b}})Liu, Lin, Hewitt, Paranjape,
  Bevilacqua, Petroni, and Liang}]{liu2023lost}
Liu, N.~F.; Lin, K.; Hewitt, J.; Paranjape, A.; Bevilacqua, M.; Petroni, F.;
  and Liang, P. 2023{\natexlab{b}}.
\newblock Lost in the Middle: How Language Models Use Long Contexts.
\newblock arXiv:2307.03172.

\bibitem[{Lu et~al.(2022)Lu, Bartolo, Moore, Riedel, and
  Stenetorp}]{DBLP:conf/acl/LuBM0S22}
Lu, Y.; Bartolo, M.; Moore, A.; Riedel, S.; and Stenetorp, P. 2022.
\newblock Fantastically Ordered Prompts and Where to Find Them: Overcoming
  Few-Shot Prompt Order Sensitivity.
\newblock In \emph{{ACL} {(1)}}, 8086--8098. Association for Computational
  Linguistics.

\bibitem[{Mao et~al.(2021)Mao, Zhu, Wang, Dai, Dong, Xiao, and
  He}]{DBLP:conf/cikm/MaoZWDDXH21}
Mao, K.; Zhu, J.; Wang, J.; Dai, Q.; Dong, Z.; Xiao, X.; and He, X. 2021.
\newblock SimpleX: {A} Simple and Strong Baseline for Collaborative Filtering.
\newblock In \emph{{CIKM}}, 1243--1252. {ACM}.

\bibitem[{OpenAI(2022)}]{chatgpt}
OpenAI. 2022.
\newblock Introducing ChatGPT.
\newblock \emph{CoRR}.

\bibitem[{Papadamou et~al.(2022)Papadamou, Zannettou, Blackburn, De~Cristofaro,
  Stringhini, and Sirivianos}]{papadamou2022just}
Papadamou, K.; Zannettou, S.; Blackburn, J.; De~Cristofaro, E.; Stringhini, G.;
  and Sirivianos, M. 2022.
\newblock “It is just a flu”: Assessing the Effect of Watch History on
  YouTube’s Pseudoscientific Video Recommendations.
\newblock In \emph{Proceedings of the international AAAI conference on web and
  social media}, volume~16, 723--734.

\bibitem[{Qin et~al.(2023)Qin, Jagerman, Hui, Zhuang, Wu, Shen, Liu, Liu,
  Metzler, Wang, and Bendersky}]{qin2023large}
Qin, Z.; Jagerman, R.; Hui, K.; Zhuang, H.; Wu, J.; Shen, J.; Liu, T.; Liu, J.;
  Metzler, D.; Wang, X.; and Bendersky, M. 2023.
\newblock Large Language Models are Effective Text Rankers with Pairwise
  Ranking Prompting.
\newblock arXiv:2306.17563.

\bibitem[{Turpin et~al.(2023)Turpin, Michael, Perez, and
  Bowman}]{unfaithful-cot}
Turpin, M.; Michael, J.; Perez, E.; and Bowman, S.~R. 2023.
\newblock Language Models Don't Always Say What They Think: Unfaithful
  Explanations in Chain-of-Thought Prompting.
\newblock \emph{CoRR}, abs/2305.04388.

\bibitem[{Wang and Lim(2023)}]{wang2023zeroshot}
Wang, L.; and Lim, E.-P. 2023.
\newblock Zero-Shot Next-Item Recommendation using Large Pretrained Language
  Models.
\newblock arXiv:2304.03153.

\bibitem[{Wang et~al.(2023{\natexlab{a}})Wang, Li, Chen, Zhu, Lin, Cao, Liu,
  Liu, and Sui}]{wang2023large}
Wang, P.; Li, L.; Chen, L.; Zhu, D.; Lin, B.; Cao, Y.; Liu, Q.; Liu, T.; and
  Sui, Z. 2023{\natexlab{a}}.
\newblock Large Language Models are not Fair Evaluators.
\newblock arXiv:2305.17926.

\bibitem[{Wang et~al.(2023{\natexlab{b}})Wang, Lin, Feng, He, and
  Chua}]{wang2023generative}
Wang, W.; Lin, X.; Feng, F.; He, X.; and Chua, T.-S. 2023{\natexlab{b}}.
\newblock Generative Recommendation: Towards Next-generation Recommender
  Paradigm.
\newblock arXiv:2304.03516.

\bibitem[{Wu et~al.(2022)Wu, Wu, Qi, Liu, Tian, Li, He, Huang, and
  Xie}]{wu2022feedrec}
Wu, C.; Wu, F.; Qi, T.; Liu, Q.; Tian, X.; Li, J.; He, W.; Huang, Y.; and Xie,
  X. 2022.
\newblock Feedrec: News feed recommendation with various user feedbacks.
\newblock In \emph{Proceedings of the ACM Web Conference 2022}, 2088--2097.

\bibitem[{Wu et~al.(2020)Wu, Qiao, Chen, Wu, Qi, Lian, Liu, Xie, Gao, Wu, and
  Zhou}]{DBLP:conf/acl/WuQCWQLLXGWZ20}
Wu, F.; Qiao, Y.; Chen, J.; Wu, C.; Qi, T.; Lian, J.; Liu, D.; Xie, X.; Gao,
  J.; Wu, W.; and Zhou, M. 2020.
\newblock {MIND:} {A} Large-scale Dataset for News Recommendation.
\newblock In \emph{{ACL}}, 3597--3606. Association for Computational
  Linguistics.

\bibitem[{Wu et~al.(2023)Wu, Zheng, Qiu, Wang, Gu, Shen, Qin, Zhu, Zhu, Liu,
  Xiong, and Chen}]{wu2023survey}
Wu, L.; Zheng, Z.; Qiu, Z.; Wang, H.; Gu, H.; Shen, T.; Qin, C.; Zhu, C.; Zhu,
  H.; Liu, Q.; Xiong, H.; and Chen, E. 2023.
\newblock A Survey on Large Language Models for Recommendation.
\newblock arXiv:2305.19860.

\bibitem[{Xu et~al.(2022)Xu, Xu, Chen, Dong, and Wen}]{DBLP:conf/cikm/Xu0CDW22}
Xu, C.; Xu, J.; Chen, X.; Dong, Z.; and Wen, J. 2022.
\newblock Dually Enhanced Propensity Score Estimation in Sequential
  Recommendation.
\newblock In \emph{{CIKM}}, 2260--2269. {ACM}.

\bibitem[{Zhang et~al.(2021)Zhang, DING, Shui, Ma, Zou, Deoras, and
  Wang}]{zhang2021language}
Zhang, Y.; DING, H.; Shui, Z.; Ma, Y.; Zou, J.; Deoras, A.; and Wang, H. 2021.
\newblock Language Models as Recommender Systems: Evaluations and Limitations.
\newblock In \emph{I (Still) Can't Believe It's Not Better! NeurIPS 2021
  Workshop}.

\bibitem[{Zhao et~al.(2021)Zhao, Wallace, Feng, Klein, and
  Singh}]{DBLP:conf/icml/ZhaoWFK021}
Zhao, Z.; Wallace, E.; Feng, S.; Klein, D.; and Singh, S. 2021.
\newblock Calibrate Before Use: Improving Few-shot Performance of Language
  Models.
\newblock In \emph{{ICML}}, volume 139 of \emph{Proceedings of Machine Learning
  Research}, 12697--12706. {PMLR}.

\end{thebibliography}

\end{document}